\documentclass[a4paper,aps,11pt,nofootinbib,showpacs]{revtex4}
\usepackage{amsmath}
\usepackage{amssymb}
\usepackage{amsfonts}
\usepackage[dvips]{graphicx}

\begin{document}
\newcommand{\del}[1]{\partial_{#1}}

\title{Dynamics of Toroidal Spiral Strings around Five-dimensional Black Holes}
\today
\hfill{OCU-PHYS 322}

\hfill{AP-GR 73}


\author{Takahisa Igata} 
\email{igata@sci.osaka-cu.ac.jp}
\author{Hideki Ishihara}
\email{ishihara@sci.osaka-cu.ac.jp}
\affiliation{%
 Department of Mathematics and Physics,
 Graduate School of Science, Osaka City University,
 Osaka 558-8585, Japan}

\begin{abstract}
We examine the separability of the Nambu-Goto equation for test strings 
in a shape of toroidal spiral in a five-dimensional Kerr-AdS black hole. 
In particular, 
for a \lq{\it Hopf loop}\rq\ string which is a special class of 
the toroidal spiral strings, 
we show the complete separation of variables occurs in two cases, 
Kerr background and Kerr-AdS background with equal angular momenta. 
We also obtain the dynamical solution for the Hopf loop around a black hole 
and for the general toroidal spiral in Minkowski background. 
\end{abstract}
\maketitle

\section{Introduction}
Recently, much attention has been focused on the study of higher-dimensional 
spacetimes in relation to ideas of AdS/CFT correspondence \cite{Maldacena:1997re, 
Gubser:1998bc, Witten:1998qj}
and braneworld \cite{ArkaniHamed:1998rs, Antoniadis:1998ig, 
Randall:1999ee, Randall:1999vf}. 
In the context of the topics, 
higher-dimensional black hole spacetimes would play crucial roles. 
In the braneworld scenarios, 
since matter fields are confined on the four-dimensional membrane 
while the gravity propagates in all dimensions, 
the observation of phenomena related to the gravity is the only way to see 
the existence of extra-dimensions. 
For this reason, physical properties in the strong gravitational field 
around higher-dimensional black holes should be illuminated 
to explore the higher-dimensional spacetime.

A typical exact solution of a higher-dimensional rotating black hole 
is a generalization of the Kerr metric found by Myers and Perry \cite{Myers:1986un}, 
and the solution was extended to the Kerr-AdS 
black hole solution \cite{Hawking:1998kw, Gibbons:2004js}. 
Furthermore, it has been found that there are several black holes which have a 
variety of the horizon topology such as black strings and black rings 
(see \cite{LivingReview} for a review). 
It has been recognized that the higher-dimensional black holes possess richer 
properties than the four-dimensional black holes.

Though there are differences in properties of black holes in four-dimensions and 
in higher dimensions, 
it is known that the Kerr-AdS black holes 
in general dimensions have a remarkable common property, 
that is, separation of variables in the geodesic 
Hamilton-Jacobi equation \cite{Carter:1968ks, Frolov:2003en, 
Frolov:2006pe,Frolov:2007nt}. 
This is because the Kerr-AdS black hole metrics in general dimensions 
admit rank-2 Killing tensor fields.

A geodesic particle plays an important role as a probe of black hole spacetime 
because it gives us information of the geometry around the black hole. 
In addition to a particle, a string is regarded as a probe of geometry 
of a higher-dimensional black hole. 
It has been also found that separation of variables in the Hamilton-Jacobi equation 
for the stationary strings occurs 
in the Kerr-AdS metric \cite{Frolov:2004qw, Kubiznak:2007ca, Ahmedov:2008pq}.

In this paper we study dynamics of a Namu-Goto string 
in five-dimensional spacetimes. 
We consider a special string that we call a 
{\it toroidal spiral string} \cite{BlancoPillado:2007iz, Igata:2009dr}, 
which has a spiral shape along a circle on a time slice. 
The toroidal spiral string is a cohomogeniety-one 
string \cite{Frolov, Ishihara:2005nu} associated with the Killing vector field 
which generates a combination of two commutable rotational isometry. 
Here we discuss dynamics of the toroidal spiral string in the Kerr-AdS background, 
and examine the separability of the Hamilton-Jacobi equation for the string.

In particular, we study \lq{\it Hopf loop}\rq\ strings, which are in a 
special class of the toroidal spirals. 
We consider five-dimensional spacetimes of which time slices are 
foliated by the $S^3$. 
It is well known that the $S^3$ is a Hopf fibration, 
{\it i.e.,} 
a twisted $S^1$ bundle over $S^2$. 
We call a closed string which lies along the Hopf fiber a Hopf loop.
We show that separation of variables occurs for the Hopf loop 
in the Kerr metric with two independent rotations, and 
in the Kerr-AdS metric with the equal rotations in five-dimensions. 
We also analyze dynamics of the Hopf loop in the latter black hole metrics, 
and general toroidal spirals in five-dimensional flat metric, 
where the separation of variables occurs. 

This paper is organized as follows. 
In Sec.\ref{sec:two}, we briefly review the formalism of a cohomogeneity-one string. 
In Sec.\ref{sec:three}, 
we introduce the toroidal spiral strings 
in the five-dimensional Kerr-AdS black hole spacetime, 
and study the separability of the Hamilton-Jacobi equation corresponding to 
the Nambu-Goto equation for the string. 
In Sec.\ref{sec:four}, We show that complete separation of variables occurs 
for the Hopf loop strings 
in two cases of the black hole geometry and discuss dynamical properties of 
the Hopf loops in the five-dimensional black holes.
We obtain solutions of general toroidal spirals explicitly 
in five-dimensional Minkowski background in Sec.\ref{sec:five}. 
Finally, we summarize results in Sec.\ref{sec:six}. 
We use the sign convention $- + + + +$ for the metric, 
and units in which $c = G = 1$. 

\section{Cohomogeneity-one String Motion}
\label{sec:two}

A test string motion is described by a two-dimensional worldsheet $\Sigma$ 
in a target spacetime $({\cal M} , g_{MN})$. The embedding of $\Sigma$ 
in ${\cal M}$ is determined by the parametric equation
\begin{eqnarray}
y^{M} = y^{M}(\zeta^a),
\end{eqnarray}
where $y^{M}$ are coordinates of ${\cal M}$ and 
$\zeta^a (\zeta^0 = \tau, \zeta^1 = \sigma)$ are coordinates of $\Sigma$. 
We assume that the dynamics of the string is governed by the Nambu-Goto action 
\begin{eqnarray}
S_{NG} = 
- \mu \int_{\Sigma}d^2\zeta \sqrt{- \gamma},
\end{eqnarray}
where parameter $\mu$ denotes the string tension and $\gamma$ is the determinant of 
the induced metric on $\Sigma$ given by
\begin{eqnarray}
\gamma_{ab} 
	= g_{MN} 
	\frac{\partial y^{M}}{\partial \zeta^{a}} 
	\frac{\partial y^{N}}{\partial \zeta^{b}}.
\end{eqnarray}

Let us suppose that the background spacetime metric $g_{MN}$ possesses Killing 
vector fields. If one of the Killing vector fields, say $\xi$, is tangent 
to $\Sigma$, we call the string a {\it cohomogeneity-one string} 
associated with the Killing vector field $\xi$ \cite{Frolov, Ishihara:2005nu}. 
A {\it stationary string} is one of the cohomogeneity-one string. 
The Killing vector field associated with the stationary string 
is timelike on $\Sigma$.

The group action of isometry generated by $\xi$ defines Killing orbits 
in ${\cal M}$. One can consider a set ${\cal O}$ of the Killing orbits as 
a quotient space of ${\cal M}$, on which the metric is naturally introduced 
by the projection tensor with respect to $\xi$,
\begin{eqnarray}
	h_{MN} = g_{MN} - \xi_{M} \xi_{N}/F,
\label{eq:reduced_metric}
\end{eqnarray}
where $F = \xi^{M} \xi_{M}$. The action for the cohomogeneity-one string 
associated with $\xi$ is reduced to the geodesic action for a curve ${\cal C}$ 
in ${\cal O}$ with respect to 
the metric $F h_{\mu\nu}$ \cite{Frolov, Ishihara:2005nu}
\begin{eqnarray}
	S = - \mu \int_{{\cal C}} \sqrt{- F h_{\mu\nu} dx^{\mu} dx^{\nu}},
\label{eq:reduced_action} 
\end{eqnarray}
where $x^{\mu}$ are coordinates on ${\cal O}$. 
Therefore, the problem for finding solution of cohomogeneity-one strings 
associated with Killing vector field $\xi$ reduces to the problem for solving 
geodesic equations in the 
$({\rm dim}{\cal M}-1)$-dimensional space $({\cal O},  F h_{\mu \nu})$.
Signature of $F h_{\mu\nu}$ is Euclidean for a timelike $\xi$, and is Lorentzian 
for a spacelike $\xi$. 
There are a lot of works dealing with the cohomogeneity-one strings in a variety 
of context \cite{Burden, deVega, Frolov, Ishihara:2005nu, Ogawa}. 

\section{Toroidal Spiral Strings in Five-dimensional Black Holes }
\label{sec:three}

We consider now the motion of a cohomogeneity-one string in a five-dimensional 
rotating black hole with a cosmological constant. The corresponding metric of 
these five-dimensional Kerr-AdS black holes in the Boyer-Lindquist type 
coordinates \cite{Gibbons:2004js} is 
\begin{eqnarray}
	ds^2
 		&=& - \frac{\Delta_{\theta} \Xi_r dt^2 }{\Xi_a \Xi_b}
 		+ \frac{2 M}{\Sigma}  \Big(\frac{\Delta_{\theta} dt}{\Xi_a \Xi_b}
 		- \nu \Big)^2
		+ \frac{\Sigma dr^2}{\Delta_r}
\nonumber\\[0.3cm]
&&
 + \frac{\Sigma d\theta^2}{\Delta_{\theta}}
 + \frac{r^2 + a^2}{\Xi_a} \sin^2\theta d\Phi^2
 + \frac{r^2 + b^2}{\Xi_b} \cos^2\theta d\Psi^2\ ,
\label{eq:metric}
\end{eqnarray}
with
\begin{eqnarray}
	\Xi_a = 1 - a^2 \lambda^2, 
\qquad
	\Xi_b = 1 - b^2 \lambda^2 ,
\qquad
	\Xi_r = 1 + \lambda^2 r^2\ ,
\end{eqnarray}
\begin{eqnarray}
	\Delta_r = \frac{(r^2 + a^2)(r^2 + b^2)(1 + \lambda^2 r^2)}{r^2} - 2 M , 
\qquad
	\Delta_{\theta} = 1 - a^2 \lambda^2 \cos^2\theta
		 - b^2 \lambda^2 \sin^2\theta\ ,
\end{eqnarray}
\begin{eqnarray}
	\nu = a \sin^2\theta \frac{d\Phi}{\Xi_a} 
		+ b \cos^2\theta  \frac{d\Psi}{\Xi_b} ,
\qquad
	\Sigma = r^2 + a^2 \cos^2\theta + b^2 \sin^2\theta,
\end{eqnarray}
where $M$ is the mass parameter, and $a$ and $b$ are two independent rotation 
parameters. The parameter $\lambda$ is connected with the cosmological 
constant $\Lambda$ as $\lambda^2=- \Lambda/6$. Note that the metric possesses 
three commutable Killing vector fields $\del{t} , \del{\Phi}$ and $\del{\Psi}$.

Now, let us concentrate on a special cohomogeneity-one string. 
We suppose that a worldsheet of a string is tangent to the Killing vector field
\begin{eqnarray}
	\xi = \del{\Phi} + \alpha \del{\Psi},
\label{eq:toroidal_Killing_vector}
\end{eqnarray}
where $\alpha$ is a constant. Here we name the string as a 
{\it toroidal spiral string} since the string has a toroidal spiral shape 
on a snap-shot. The constant $\alpha$ is interpreted as a ratio of winding number 
in the directions of two azimuthal angles $\Psi$ and $\Phi$. 
If $\alpha$ takes a rational number, the toroidal spiral string is closed.

We make the coordinate transformation $(\Phi , \Psi) \longmapsto  (\phi , \psi)$, 
which is defined as
\begin{alignat}{8}
	\phi &=&  \Phi,& &\qquad&
	\psi &=&  \Psi& - \alpha  \Phi,
\label{eq:coordinate_transformation}
\end{alignat}
so that $\xi=\del{\phi}$. 
Hence the quotient space ${\cal O}$ with respect to $\xi$ is covered 
by the coordinate system $(t , r , \theta , \psi)$. 
The motion of the toroidal spiral string 
is regarded as a geodesic motion in  ${\cal O}$ 
with the metric $F h_{\mu\nu}$. 
The action (\ref{eq:reduced_action}) for the geodesics 
is equivalent to the action
\begin{eqnarray}
	S = - \mu \int d\tau 
	\Big(
	\frac{1}{2 N} F h_{\mu\nu} \frac{dx^{\mu}}{d\tau} \frac{dx^{\nu}}{d\tau} 
		- \frac{N}{2}
	\Big),
\end{eqnarray}
where $N$, an arbitrary function of $\tau$, denotes a Lagrange multiplier 
which is related to the reparametrization invariance of the curve.

In order to study the geodesics, it is convenient to use the Hamilton-Jacobi method. 
The corresponding Hamilton-Jacobi equation is
\begin{eqnarray}
\frac{\partial S}{\partial \tau}
 + \frac{N}{2} 
\frac{h^{\mu\nu}}{F} 
\frac{\partial S}{\partial x^{\mu}} 
\frac{\partial S}{\partial x^{\nu}}
 = 0,
\label{eq:HJ_equation}
\end{eqnarray}
where $S$ denotes Hamilton's principal function and $F$ is given by
\begin{eqnarray}
	F = \frac{2 M}{\Sigma} \Big(\frac{a \sin^2\theta}{\Xi_a}
 		+ \alpha \frac{b \cos^2\theta}{\Xi_b}\Big)^2 
	+ \frac{(r^2 + a^2) \sin^2\theta}{\Xi_a}
 	+ \alpha^2 \frac{(r^2 + b^2) \cos^2\theta}{\Xi_b} . 
\end{eqnarray}
The contravariant components of the projection metric (\ref{eq:reduced_metric}) 
are given by
\begin{eqnarray}
 	h^{tt}
	&& = 
	- \frac{\Sigma \Xi_a \Xi_b (r^2 + a^2)(r^2 + b^2) + 2 M X}
	{r^2 \Sigma \Delta_{\theta} \Delta_{r}},
\\[0.2cm]
 	h^{\psi\psi}
	&& = \frac{1}{r^2 \Sigma \Delta_{r} \cos^2\theta \sin^2\theta}
	\left[F \Xi_a \Xi_b \Xi_r \Sigma
	 - 2 M \Delta_{\theta} 
	\Big[\frac{(r^2 + a^2) \sin^2\theta}{\Xi_a}
	 + \alpha^2 \frac{(r^2 + b^2) \cos^2\theta}{\Xi_b}\Big]
	\right],
\\[0.2cm]
 	h^{t\psi}
	&& = \frac{2 M Y}
	{r^2 \Sigma \Delta_{r}}, 
\qquad 
	h^{rr}
	 = \frac{\Delta_r}{\Sigma}, 
\qquad 
	h^{\theta\theta}
	 = \frac{\Delta_{\theta}}{\Sigma},
\end{eqnarray}
where
\begin{eqnarray}
	X = a^2 \Xi_b (r^2 + b^2)\sin^2\theta
 	+ b^2 \Xi_a (r^2 + a^2)\cos^2\theta,
\qquad
	Y = - b (r^2 + a^2)
 		+ \alpha a (r^2 + b^2).
\end{eqnarray}

Since the metric $F h_{\mu\nu}$ admits the two Killing vector fields 
$\del{t}$ and $\del{\psi}$, the momenta $p_t$ and $p_{\psi}$ take constant values, 
say $- E$ and $L$, respectively. 
Since $t$ and $\psi$ are cyclic coordinates, we can obtain a 
complete solution of the equation (\ref{eq:HJ_equation}) of the form
\begin{eqnarray}
	S =  \frac12 \mu^2 \chi - E t + L \psi + S_{r\theta},
\end{eqnarray}
where $S_{r\theta}$ is a function of $r$ and $\theta$, and $\chi$ 
is function of parameter $\tau$ such that $\dot \chi  = N$. 
Then, the system reduces to a particle system in two-dimensions described by 
$S_{r\theta}(r, \theta)$. 
Here, we assume complete separation of variables of $S_{r\theta}$ in the form
\begin{eqnarray}
	S_{r\theta}(r, \theta) = S_r(r) + S_{\theta}(\theta),
\end{eqnarray}
where $S_r$ and $S_{\theta}$ are functions of $r$ and $\theta$, respectively. 

Substituting the expression 
for Hamilton's principal function into the Hamilton-Jacobi equation, 
we can write down it in the form
\begin{eqnarray}
&&\bigg[\frac{(r^2 + a^2)(r^2 + b^2)\Xi_a \Xi_b 
- 2 M (r^2 + a^2 b^2 \lambda^2)}{r^2 \Delta_r}
 + \frac{2 (1 - a^2 \lambda^2) (1 - b^2 \lambda^2)}{\lambda^2 (a^2 + b^2) 
+ \lambda^2 (a^2 - b^2) \cos2\theta - 2}\bigg] \frac{E^2}{\lambda^2}\nonumber
\\[0.2cm]
&& + 
\bigg[
4 a b M \alpha \Xi_r
 + (\alpha^2 - 1)(b^2 - a^2) r^2 \Xi_r
 + a^2 b^2 \Xi_r 
\Big[
(\alpha^2 - 1)(1 + \lambda^2 (a^2 - b^2)) - 2 \alpha^2
\Big]\nonumber
\\ [0.2cm]
&& + a^2 
(1 + \alpha^2 r^2 \lambda^2)(a^2 \Xi_r - 2 M)
 + b^2 
(\alpha^2 + r^2 \lambda^2)(b^2 \Xi_r - 2 M)
 + \Big(
\frac{\alpha^2 \Xi_a}{\sin^2\theta} + \frac{\Xi_b}{\cos^2 \theta}
\Big)
\bigg] L^2\nonumber
\\ [0.2cm]
	&& + \frac{4 M Y}{r^2 \Delta_r} E L
 	+ \Delta_r \Big(\frac{d S_r}{d r}\Big)^2
 	+ \Delta_{\theta}\Big(\frac{d S_{\theta}}{d \theta}\Big)^2
 	= -\mu^2 F \Sigma.
\label{eq:HJeqinKerrAdS}
\end{eqnarray}
We find that the separability of variables $r$ and $\theta$ in this equation 
crucially depends only on the term in the right-hand-side. 
The explicit form of the term is given by
\begin{eqnarray}
	\mu^2 F \Sigma
	&=& \mu^2 (r^2 + a^2 \cos^2\theta + b^2 \sin^2\theta)
	 \Big[\frac{(r^2 + b^2)\alpha^2 \cos^2\theta}{\Xi_b} 
		+ \frac{(r^2 + a^2) \sin^2\theta}{\Xi_a} \Big]
\nonumber\\ 
	&& + 2 M \mu^2 
	\Big(
	\frac{\alpha b \cos^2\theta}{\Xi_b} 
	+ \frac{a \sin^2\theta}{\Xi_a}\Big)^2.
\label{eq:mass_term}
\end{eqnarray}
The complete separation of variables in the Hamilton-Jacobi equation depends 
on the parameter $\alpha$ which describes winding ratio of the string, 
and parameters of the background geometry. 
In a particular case of toroidal spirals with $\alpha^2=1$, 
the separation of variables occurs in two cases of the black hole geometry. 
We see this fact in the following section.

\section{Hopf loops}
\label{sec:four}

In this section, we concentrate on the case of $\alpha^2= 1$. 
As is shown below, the toroidal spiral string of this case lies 
along a fiber of the Hopf fibration of $S^3$, 
then we call the string with $\alpha^2=1$ Hopf loop string.

Let us suppose a timeslice of the spacetime with the metric \eqref{eq:metric}. 
The time slice is foliated by $r=const.$ surfaces, which have the topology 
of $S^3$, with the metric  
\begin{eqnarray}
	ds_{S^3}^2
		=&& g_{\theta\theta}d\theta^2
 		+ g_{\Phi\Phi}d\Phi^2 +g_{\Psi\Psi} d\Psi^2+ 2 g_{\Phi\Psi} d\Phi d\Psi
\nonumber\\[2mm]
 = &&  \frac{\Sigma }{\Delta_{\theta}}d\theta^2
 + \left( \frac{r^2 + a^2}{\Xi_a}
	+ \frac{2Ma^2}{\Sigma \Xi_a^2} \sin^2\theta \right)\sin^2\theta d\Phi^2
\cr
&&
\hspace{1.2cm} + \left( \frac{r^2 + b^2}{\Xi_b}
	+ \frac{2Mb^2}{\Sigma\Xi_b^2} \cos^2\theta \right) \cos^2\theta d\Psi^2
	+\frac{4 Mab}{\Sigma \Xi_a \Xi_b}\sin^2\theta  \cos^2\theta {d\Phi}{d\Psi} .
\label{eq:S3_metric}
\end{eqnarray}
The metric describes $S^3$ that is deformed from the round metric 
by the continuous parameters $M, a, b$ and $\lambda$. 
The metric \eqref{eq:S3_metric} is rewritten as 
\begin{eqnarray}
	ds^2_{S^3}
	=\frac14 \Big[
	g_{\theta\theta}d\theta_E^2
 		+ \frac{4}{F}(g_{\Phi\Phi}g_{\Psi\Psi} - g_{\Phi\Psi}^2)d\phi_E^2 \Big]
 		+ \frac{F}{4}\Big[
			d\psi_E + \frac{1}{F}(g_{\Psi\Psi} - g_{\Phi\Phi})d\phi_E
		\Big]^2 ,
\label{eq:Hopf_fibration}
\end{eqnarray}
where 
\begin{eqnarray}
\theta=\frac{\theta_E}{2},
\quad
\Phi=\frac12 (\psi_E - \phi_E),
\quad
\Psi=\frac12 (\psi_E + \phi_E), 
\end{eqnarray}
and
\begin{eqnarray}
	F = g_{\Phi\Phi} + g_{\Psi\Psi}  + 2g_{\Phi\Psi} . 
\end{eqnarray}
The metric \eqref{eq:Hopf_fibration} provides the scheme of the Hopf fibration 
of $S^3$, where the projection along the integral curves of the Killing vector field 
$\partial_{\psi_E}$ defines a map from $S^3$ to $S^2$. 
The first term in the metric \eqref{eq:Hopf_fibration} is the 
metric on the $S^2$ base space, and the second term is the metric on the $S^1$ fiber 
of the Hopf fibration of $S^3$. 
If $a = b$, the base space is round $S^2$, otherwise the base space $S^2$ is 
deformed by the rotation in the direction of $\phi_E$. 

On each timeslice, the toroidal spiral with $\alpha=1$, 
which is associated with the Killing vector field
$\xi=\partial_\Phi+\partial_\Psi$, 
lies on a fiber of the Hopf fibration of $S^3$ because 
$\xi=2\partial_{\psi_E}$, {\it i.e.}, the tangent vector field of the string 
generates the $S^1$ fiber of $S^3$. 
Therefore, we refer the toroidal spiral string with $\alpha=1$ as a {\it Hopf loop}.
The toroidal spiral string with $\alpha=-1$ is also the Hopf loop 
which is obtained by the coordinate reflection $\Psi \mapsto - \Psi$.

When we consider the Hopf loop around the black hole, we see 
that the Hamilton-Jacobi equation written in the Boyer-Lindquist coordinates can be 
separable for two cases: 
(i) 
the vanishing cosmological constant {\it i.e.}, the background is a Kerr black hole, 
(ii) 
the black hole with non-zero cosmological constant and two equal angular momenta. 
In what follows we discuss the separability in the two cases separately. 

\subsection{Five-dimensional Kerr background}

For a Hopf loop in the Kerr background, {\it i.e.}, vanishing cosmological constant, 
the equation (\ref{eq:HJeqinKerrAdS}) 
is separated into two independent equations of $r$ and $\theta$ 
because $\mu^2 F \Sigma$ consists of three terms as 
\begin{eqnarray}
\mu^2 F \Sigma 
	= \mu^2 (r^2 + a^2)(r^2 + b^2)
 + \mu^2 (a^2 - b^2)^2 \cos^2\theta \sin^2\theta
 + 2 M \mu^2 (a \sin^2\theta + b \cos^2\theta)^2, 
\end{eqnarray}
where the first term depends only on $r$, and the second and third on $\theta$. 
Therefore the separated equations are given by
\begin{eqnarray}
&&\Delta_r 
\Big(
\frac{d S_r}{d r}
\Big)^2
 - \frac{(r^2 + a^2)(r^2 + b^2)(r^2 + 2 M)
 - 2 M r^4}
{(r^2 + a^2)(r^2 + b^2) - 2 M r^2} E^2
 + \frac{[(a^2 - b^2)^2 - 2 M (a - b)^2] L^2}
{r^2 \Delta_r}
\nonumber\\[0.1cm]
&&\quad + \frac{4 M[- b (r^2 + a^2) + a (r^2 + b^2)] E L}
{r^2 \Delta_r}
 + \mu^2 (r^2 + a^2)(r^2 +b^2)
 = - K,
\end{eqnarray}
and
\begin{eqnarray}
&&\Delta_{\theta}
\Big(\frac{d S_{\theta}}{d \theta} \Big)^2
 - (a^2 \cos^2\theta + b^2 \sin^2\theta) E^2
 + \Big(\frac{1}{\cos^2\theta} + \frac{1}{\sin^2\theta} \Big) L^2
\nonumber\\[0.1cm]
&&\quad + \mu^2 
\Big[ 2 M (a \sin^2\theta + b \cos^2\theta)^2
 + (a^2 - b^2)^2 \sin^2\theta \cos^2\theta \Big]
 = K,
\label{eq:separatedEqforKerr}
\end{eqnarray}
where $K$ is a separation constant. 
The complete separability implies that the metric $F h_{\mu\nu}$ admits a 
Killing tensor field $K^{\mu\nu}$ which gives the quadratic 
constant of motion $K = K^{\mu\nu} p_{\mu} p_{\nu}$. 
Using equation (\ref{eq:separatedEqforKerr}), one obtains the irreducible 
Killing tensor field 
\begin{eqnarray}
K^{\mu\nu}
&& = \Delta_{\theta} 
\delta_{\theta}^\mu \delta_{\theta}^\nu
 - (a^2 \cos^2\theta + b^2 \sin^2\theta) 
\delta_{t}^\mu \delta_{t}^\nu
 + 
\Big(
\frac{1}{\cos^2\theta}
 + \frac{1}{\sin^2\theta}
\Big)
 \delta_{\psi}^\mu \delta_{\psi}^\nu
\nonumber\\[0.1cm]
&& - \Big[2 M (a \sin^2\theta + b \cos^2\theta)^2
 + (a^2 - b^2)^2 \sin^2\theta \cos^2\theta \Big] h^{\mu\nu}.
\end{eqnarray}
Therefore the motion of the Hopf loop in Kerr background can be integrable.

\subsection{Five-dimensional Kerr-AdS with equal angular momenta}

For a Hopf loop with $\alpha=\pm1$ in Kerr-AdS background 
with $a = \pm b$, 
the expression (\ref{eq:HJeqinKerrAdS}) is also separated into two independent 
equations of $r$ and $\theta$ because $\mu^2 F \Sigma$ is simply the function 
of $r$ in the form
\begin{eqnarray}
\mu^2 F \Sigma
 = \frac{\mu^2}{\Xi_a^2} \Big[
\Xi_a (r^2 + a^2)^2 + 2 M a^2
\Big].
\end{eqnarray}
Then, we obtain that
\begin{eqnarray}
&&\Delta_r 
\Big(
\frac{d S_r}{d r}
\Big)^2
 - \frac{\Xi_a (r^2 + a^2)^3 + 2 M a^2 (r^2 + a^2)}
{r^2 \Delta_r} E^2
 + \frac{\mu^2 [\Xi_a (r^2 + a^2)^2 + 2 M a^2]}
{\Xi_a^2}
 = - K,
\label{eq:HJ_r}
\end{eqnarray}
and
\begin{eqnarray}
(1-a^2 \lambda^2) \Big[\Big(
\frac{d S_{\theta}}{d \theta}\Big)^2
 &+& \Big(\frac{1}{\cos^2\theta}
 + \frac{1}{\sin^2\theta} \Big) L^2\Big]
 = K.
\label{eq:HJ_theta}
\end{eqnarray}
Thus, the Hopf loop motions in the background of the Kerr-AdS black hole 
with two equal angular momenta are completely integrable. By using the 
above expression, one can read the form of the Killing tensor field
\begin{eqnarray}
K^{\mu\nu}
 = (1-a^2 \lambda^2) 
\Big[
\delta_{\theta}^{\mu} 
\delta_{\theta}^{\nu}
 + 
\Big(
\frac{1}{\cos^2\theta}
 + \frac{1}{\sin^2\theta}
\Big) 
\delta_{\psi}^{\mu} 
\delta_{\psi}^{\nu}
\Big]. 
\label{eq:reducibleKT}
\end{eqnarray}

Here, let us see the metric $Fh_{\mu\nu}$ on the quotient space $\cal O$. 
As is mentioned above, the metric of the base space in \eqref{eq:Hopf_fibration} 
becomes round $S^2$ in the case $a =b$, and $F$, the norm of $\xi$, is a function 
of $r$. 
Then, the metric $Fh_{\mu\nu}$ admits SO(3) isometry group. 
Therefore, we can restrict our attention to study geodesics confined 
in the equatorial plane, 
{\it i.e.}, $\theta_E=2\theta=\pi/2$, without loss of generality\footnote{
Since the symmetry of the quotient space $({\cal O}, F h_{\mu \nu})$ are 
enhanced in the case $a=b$, the Killing tensor field \eqref{eq:reducibleKT} is 
reducible. 
}. 

Thus, setting 
\begin{eqnarray}
\frac{dS_\theta}{d\theta}=0 \quad{\rm and}\quad \theta=\pi/4
\end{eqnarray}
in \eqref{eq:HJ_theta},  we have 
\begin{eqnarray}
	K=4L^2 \Xi_a .  
\end{eqnarray}
Then the Hamilton-Jacobi equation \eqref{eq:HJ_r} becomes 
\begin{eqnarray}
\Delta_r \Big(\frac{dS_r}{dr}\Big)^2
-\frac{a^2 \Delta_a}{r^2 \Delta_r}(r^2+a^2)E^2
+\frac{\mu^2 a^2 \Delta_a}{\Xi_a^2}+4 L^2 \Xi_a =0, 
\end{eqnarray}
where
\begin{eqnarray}
	a^2 \Delta_a := (r^2+a^2)^2\Xi_a + 2 M a^2.
\end{eqnarray}

We can obtain Hamilton's principal function in the form
\begin{eqnarray}
  S	= \frac12 \mu^2 \chi - E t + L \psi + \sigma_r \int dr \sqrt{\Theta_r},
\end{eqnarray}
where $\Theta_r$ is given by 
\begin{eqnarray}
 \Theta_r
	= \frac{1}{\Delta_r}
	\Big[\frac{a^2 \Delta_a}{r^2 \Delta_r}(r^2+a^2)E^2
	-\frac{\mu^2 a^2 \Delta_a}{\Xi_a^2} -4 L^2 \Xi_a \Big],
\end{eqnarray}
and $\sigma_r=\pm 1$.
By setting the derivatives of $S$ with respect to $\mu , E$ and $L$ equal to zero, 
we get a solution for the Hamilton-Jacobi equation in the following form
\begin{eqnarray}
&&\chi = \sigma_r \int dr \frac{a^2 \Delta_a}{\Delta_r \Xi_a^2 \sqrt{\Theta_r}},
\\
&&t=
\sigma_r \int dr \frac{a^2 \Delta_a (r^2 + a^2)E}{r^2 \Delta_r^2 \sqrt{\Theta_r}},
\\
&&\psi = \sigma_r \int dr \frac{4 L \Xi_a}{\Delta_r \sqrt{\Theta_r}}.
\end{eqnarray}
By differentiating these equations with respect to $\tau$, one obtains 
the first order differential equations
\begin{eqnarray}
&&\frac1N \dot r 
= \sigma_r \frac{\Delta_r \Xi_a^2}{a^2 \Delta_a}\sqrt{\Theta_r},
\label{eq:EOM_r}
\\
&&\frac1N \dot t 
= \frac{(r^2+a^2)\Xi_a^2 E}{r^2 \Delta_r},
\\
&&\frac1N \dot \psi
= \frac{4 L \Xi_a^3}{a^2 \Delta_a}.
\end{eqnarray}

\begin{figure}[htbp]
 \begin{center}
 \includegraphics[width=150mm]{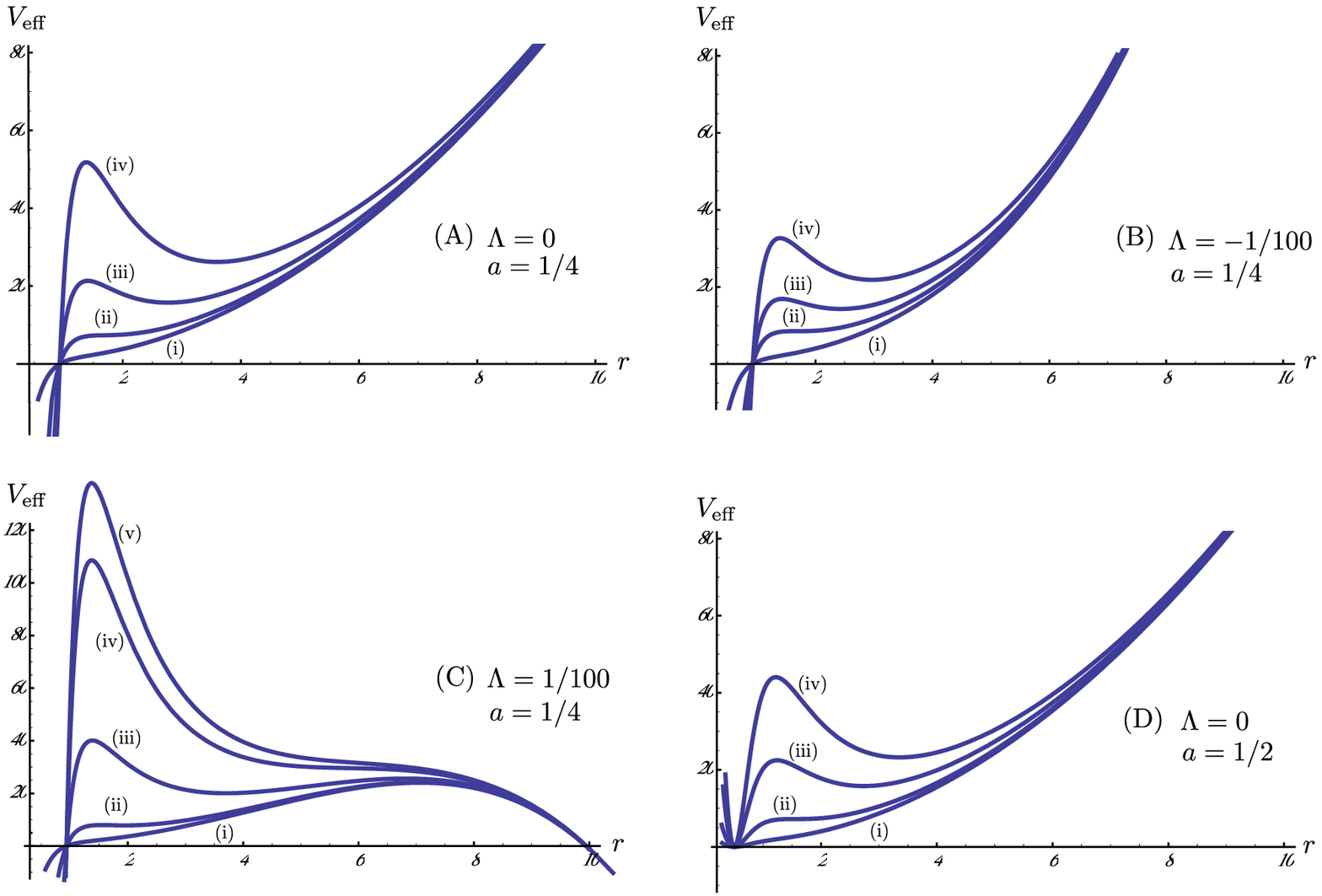}
 \end{center}
 \caption{The effective potential for radial motion of a Hopf loop 
in the five-dimensional Kerr-AdS metric with equal angular momenta. 
The mass parameter is chosen as $2M = 1 $. 
The parameters in the panels are the following. \\ 
(A) $\Lambda=0$, $a=1/4$; 
(i)$L^2 = 1$,
(ii)$L^2 = 6$, 
(iii)$L^2 = 20$, 
(iv)$L^2 = 50$. \\
(B) $\Lambda=-1/100$, $a=1/4$; 
(i)$L^2 = 1$,
(ii)$L^2 = 6.9$, 
(iii)$L^2 = 15$, 
(iv)$L^2 = 30$. \\
(C) $\Lambda=1/100$, $a=1/4$; 
(i)$L^2 = 1$,
(ii)$L^2 = 7$, 
(iii)$L^2 = 40$, 
(iv)$L^2 = 110$, 
(v)$L^2 = 140$. \\
(D) $\Lambda=0$, $a=1/2$; 
(i)$L^2 = 1$,
(ii)$L^2 = 5.7$, 
(iii)$L^2 = 20$, 
(iv)$L^2 = 40$. 
}
\label{fig:effective_potential}
\end{figure}

If we choose the Lagrange multiplier $N$ as
\begin{eqnarray}
N^2 = \frac{a^2 \Delta_a r^2}{\Xi_a^4 (r^2 + a^2)},
\end{eqnarray}
the radial motion is determined by
\begin{align}
	\dot r^2+V_{{\rm eff}}=E^2, 
\end{align}
where the effective potential $V_{{\rm eff}}$ is given by
\begin{eqnarray}
	V_{{\rm eff}} = \frac{\mu^2 r^2 \Delta_r}{(r^2 + a^2)\Xi_a^2}
	 + \frac{4 r^2 \Delta_r \Xi_a L^2}{a^2 \Delta_a (r^2 + a^2)}.
\end{eqnarray}
The effective potential shapes are given in FIG.\ref{fig:effective_potential}.

FIG.\ref{fig:effective_potential} shows that  
the radial motion of the Hopf loop is classified into two types, 
bounded or unbounded.
The existence of bounded orbits for the Hopf loop is 
analogous to the case of a geodesic particle around a four-dimensional black hole. 
We note that there is no bounded orbit around the five-dimensional 
Kerr black hole \cite{Frolov:2003en}. 
Stationary Hopf loop solution exists at the local minimum of $V_{\rm eff}$. 
By the effect of the gravitational force, there exists a critical radius for 
each black hole such that no stable Hopf loop inside the radius, namely, 
the innermost stable orbit exists. 
In addition, in the case of $\Lambda > 0$, Hopf loops can grow up to infinite 
radius by the de Sitter expansion, 
and the outermost stable orbit exists. 
In the panel (D) of FIG.\ref{fig:effective_potential}, 
we see that the radius of innermost stable orbit 
does not touch the degenerate horizon of the extremal black hole. 
This property is different from the case of a geodesic particle 
in four-dimensional extremal black hole.

\section{Dynamics of Toroidal Spiral Strings 
in Five-dimensional Minkowski background}
\label{sec:five}

In this section, we consider general toroidal spiral strings 
in Minkowski background. 
Stationary solutions of toroidal spirals in five-dimensional flat 
spacetime are studied in Ref.\cite{BlancoPillado:2007iz, Igata:2009dr}. 
Here, we investigate the dynamical motions of the toroidal spirals. 
Already we know that the Hamilton-Jacobi equation is not separable for 
$\alpha^2\neq 1$ in the $(r, \theta)$ coordinates 
even in the case of Minkowski background. 
However, we find that the Hamilton-Jacobi equation is separable 
for general $\alpha$ in the Minkowski background by using the new coordinates  
defined by 
\begin{eqnarray}
	\rho = r \sin \theta, \qquad 
	\zeta = r \cos \theta.
\label{eq:new_coordinates}
\end{eqnarray}
Furthermore, $(\rho, \zeta)$ coordinates make the physical description 
clear.
The Minkowski metric in the $(\rho, \zeta)$ coordinates becomes
\begin{eqnarray}
	ds^2 = - dt^2 + d\rho^2 + \rho^2 d\Phi^2 + d\zeta^2 + \zeta^2 d\Psi^2 . 
\end{eqnarray}
Two couples of coordinates $(\rho ,\Phi)$ and $(\zeta , \Psi)$ cover two 
independent flat planes.

Now we can apply the Hamilton-Jacobi method to the dynamical system 
in the coordinates \eqref{eq:new_coordinates}. 
We transform the angular coordinates $(\Phi , \Psi)$ into $(\phi ,\psi)$ 
by (\ref{eq:coordinate_transformation}). 
Then in the coordinates $(t, \rho, \zeta, \psi)$ on the quotient space ${\cal O}$ 
with respect to $\xi$ in \eqref{eq:toroidal_Killing_vector}, 
the contravariant components of the effective metric are given by
\begin{eqnarray}
 \frac{h^{\mu\nu}}{F}
 = \frac{1}{F} 
	\big(- \delta_{t}^\mu \delta_{t}^\nu + \delta_{\rho}^\mu \delta_{\rho}^\nu
 		+ \delta_{\zeta}^\mu \delta_{\zeta}^\nu\big)
 	+ \frac{1}{\rho^2 \zeta^2} \delta_{\psi}^\mu \delta_{\psi}^\nu ,
\end{eqnarray}
where $F = \rho^2 + \alpha^2 \zeta^2$. In the same way as in the previous section, 
Hamilton's principal function is assumed to be
\begin{eqnarray}
	S = \frac12 \mu^2 \chi - E t + L \psi + S_\rho + S_\zeta,
\end{eqnarray}
where $S_\rho$ and $S_\zeta$ are functions of $\rho$ and $\zeta$, respectively. 
Substituting this expression into the Hamilton-Jacobi equation, one obtains 
\begin{eqnarray}
- E^2  + \Big(\frac{1}{\zeta^2} + \frac{\alpha^2}{\rho^2}\Big) L^2
 + \Big(\frac{dS_{\rho}}{d\rho}\Big)^2 + \Big(\frac{dS_{\zeta}}{d\zeta}\Big)^2
 + \mu^2 (\rho^2 + \alpha^2 \zeta^2)
 = 0.
\label{eq:HJeqinRZcoord_flat}
\end{eqnarray}
For all $\alpha$, the separability of the above equation is manifest.\footnote{
Note that in the Schwarzschild background, 
the separation of variables $\rho$ and $\zeta$ does not occur even for Hopf loops. }
We solve the Hamilton-Jacobi equation (\ref{eq:HJeqinRZcoord_flat}) 
for the toroidal spiral strings in Minkowski background. 
Introducing a separation constant $K$, we have 
\begin{eqnarray}
	&&\Big(\frac{dS_\rho}{d\rho}\Big)^2
	 + \mu^2 \rho^2  + \frac{\alpha^2 L^2}{\rho^2} - \frac{E^2}{2}
	 = K,
\\
	&&\Big(\frac{dS_\zeta}{d\zeta}\Big)^2
 		+ \mu^2 \alpha^2 \zeta^2 + \frac{L^2}{\zeta^2} - \frac{E^2}{2}
 		= - K . 
\end{eqnarray}
Hamilton's principal function is obtained as
\begin{eqnarray}
S
 = \frac12 \mu^2 \chi
 - E t
 + L \psi
 + \sigma_\rho \int d\rho \sqrt{\Theta_\rho}
 + \sigma_\zeta \int d\zeta \sqrt{\Theta_\zeta}.
\end{eqnarray}
where
\begin{eqnarray}
	\Theta_{\rho} &=&  
K
 - \mu^2 \rho^2
 - \frac{\alpha^2 L^2}{\rho^2}
 + \frac{E^2}{2},
\label{eq:Theta_rho}
\\
	\Theta_\zeta
 	&=& 
- K
 - \mu^2 \alpha^2 \zeta^2
 - \frac{L^2}{\zeta^2}
 + \frac{E^2}{2},
\label{eq:Theta_zeta}
\end{eqnarray}
and $\sigma_\rho$ and $\sigma_\zeta$ are sign functions. 
By differentiating $S$ with respect to $\mu,\ E,\ L$ and $K$, we obtain the solution 
of the Hamilton-Jacobi equations (\ref{eq:HJeqinRZcoord_flat}) as 
\begin{eqnarray}
&&\chi = 
  \sigma_\rho \int d\rho \frac{\rho^2}{\sqrt{\Theta_\rho}}
+ \sigma_\zeta \int d\zeta \frac{\alpha^2 \zeta^2}{\sqrt{\Theta_\zeta}},
\\[0.2cm]
&&t
 = 
\sigma_\rho \int d\rho \frac{E}{\sqrt{\Theta_\rho}},
\\[0.2cm]
&&\psi
 = L
\Big[
\sigma_\rho \int d\rho \frac{\alpha^2}{\rho^2 \sqrt{\Theta_\rho}}
 + \sigma_\zeta \int \frac{d\zeta}{\zeta^2 \sqrt{\Theta_\zeta}}
\Big],
\\[0.2cm]
&&\sigma_\rho \int \frac{d\rho}{\sqrt{\Theta_{\rho}}}
 = \sigma_\zeta \int \frac{d\zeta}{\sqrt{\Theta_\zeta}}.
\end{eqnarray}
Often, it is convenient to rewrite these equations in the form of 
the first-order differential equations
\begin{eqnarray}
&&\frac{1}{N} \dot t = \frac{E}{F},
\\
&&\frac{1}{N} \dot \psi = \frac{L}{\rho^2 \zeta^2},
\\
&&\frac{1}{N} \dot \rho = \sigma_\rho \frac{1}{F} \sqrt{\Theta_\rho},
\label{eq:ODE_rho}
\\
&&\frac{1}{N} \dot \zeta = \sigma_\zeta \frac{1}{F} \sqrt{\Theta_\zeta},
\label{eq:ODE_zeta}
\end{eqnarray}
where overdot stand for differentiation with respect to $\tau$. 

If we choose the Lagrange multiplier $N$ as
\begin{eqnarray}
	N = F = \rho^2 + \alpha^2 \zeta^2,
\end{eqnarray}
then \eqref{eq:ODE_rho} and \eqref{eq:ODE_zeta} become 
\begin{eqnarray}
&&\dot \rho^2
 + \rho^2
 + \frac{\alpha^2 L^2}{\rho^2}
 - \Big( \frac{E^2}{2} + K\Big)
 = 0,
\label{eq:EOM_rho}
\\[0.2cm]
&&\dot \zeta^2
 + \alpha \zeta^2
 + \frac{L^2}{\zeta^2}
 - \Big(\frac{E^2}{2} - K \Big)
 = 0.
\label{eq:EOM_zeta}
\end{eqnarray}
At this point, we put $\mu = 1$ since it can be absorbed by the parameters $K, L$, 
and $E$. 

We can obtain the general solution explicitly in the form
\begin{eqnarray}
\rho^2
 &=& \frac{\rho^2_{\rm max}
 - \rho^2_{\rm min}}{2} \cos(2 \tau + \delta_\rho)
 + \frac{\rho^2_{\rm max} + \rho^2_{\rm min}}{2},
\label{eq:rho_solution}
\\
\zeta^2
 &=& \frac{\zeta^2_{\rm max}
 - \zeta^2_{\rm min}}{2} \cos(2 \alpha \tau + \delta_\zeta)
 + \frac{\zeta^2_{\rm max} + \zeta^2_{\rm min}}{2},
\label{eq:zeta_solution}
\end{eqnarray}
where the constants $\delta_\rho$ and $\delta_\zeta$ are initial phases 
of $\rho$ and $\zeta$, respectively. 
It is clear that
$\rho_{{\rm min}}^2 \leq \rho^2\leq \rho_{{\rm max}}^2$ 
and 
$\zeta_{{\rm min}}^2 \leq \zeta^2\leq \zeta_{{\rm max}}^2$ 
where the maximum and minimum values of $\rho$ and $\zeta$ are give by 
\begin{eqnarray}
	\rho_{{\rm max}}^2 &=& \frac{1}{4} 
	\Big[E^2 + 2 K + \sqrt{(E^2 + 2 K + 4 \alpha L)(E^2 + 2 K - 4 \alpha L)}
\Big],
\\
\rho_{{\rm min}}^2
 &=& \frac{1}{4} 
\Big[
E^2 + 2 K
 - \sqrt{(E^2 + 2 K + 4 \alpha L)(E^2 + 2 K - 4 \alpha L)}
\Big],
\\
\zeta_{{\rm max}}^2
 &=& \frac{1}{4\alpha^2} 
\Big[
E^2 - 2 K
 + \sqrt{(E^2 - 2 K + 4 \alpha L)(E^2 - 2 K - 4 \alpha L)}
\Big],
\\
\zeta_{{\rm min}}^2
 &=& \frac{1}{4\alpha^2} \Big[E^2 - 2 K
 - \sqrt{(E^2 - 2 K + 4 \alpha L)(E^2 - 2 K - 4 \alpha L)}
\Big].
\end{eqnarray}
In the case of $\rho_{{\rm max}}=\rho_{{\rm min}}$ 
and $\zeta_{{\rm max}}=\zeta_{{\rm min}}$ ,{\it i.e.,} $K=0, E^2=4\alpha L$,
we have the stationary solution 
$\rho=\rho_0=const.$ and $\zeta=\zeta_0=const.$ where 
\begin{eqnarray}
\rho_0 = \sqrt{\alpha L},
\qquad
\zeta_0 = \sqrt{\frac{L}{\alpha}}.
\end{eqnarray}
Properties of these solutions were discussed 
in Ref.\cite{BlancoPillado:2007iz,Igata:2009dr}. 

The general solution (\ref{eq:rho_solution}) and (\ref{eq:zeta_solution}) describe 
harmonic oscillations of $\rho^2$ and $\zeta^2$ in time $t=E\tau$.
The frequency of $\rho^2$ and $\zeta^2$ are given by $2/E$ and $2\alpha/E$ 
then they draw a Lissajous figure in the $\rho^2$-$\zeta^2$ plane. 
If $\alpha$ is a rational number, 
the toroidal spiral string is closed on a snapshot, 
the Lissajous figure becomes a closed curve {\it i.e.}, 
the motion of the toroidal spiral string become periodic. 
On the other hand an open toroidal spiral string, irrational $\alpha$, 
has ergode-like motion. 

\section{Conclusion}
\label{sec:six}

In this paper, 
we have studied the separability of the Nambu-Goto equations 
for toroidal spiral strings, 
whose worldsheet is tangent to a Killing vector field, say $\xi$, 
that is a linear combination of two commutable rotational Killing vector fields 
in the five-dimensional Kerr-AdS black holes. 
Dynamics of the toroidal spiral string associated with $\xi$ is determined 
by geodesics in the quotient space with respect to $\xi$ 
with the effective metric which is given by the projection tensor 
weighted by the norm of $\xi$. 
We have used the Hamilton-Jacobi method to solve geodesic equations, 
and have shown that the Hamilton-Jacobi equation in the Boyer-Lindquist coordinates 
admits the separation of variables for the Hopf loop strings, 
in the two black hole spacetimes: 
the Kerr background and the Kerr-AdS background with equal angular momenta. 
The Hopf loop string is a special case of the toroidal spiral strings 
which lie along a fiber of the Hopf fibration of $S^3$ 
which foliates spacial sections of the target spacetime. 
The separability is owing to the existence of a rank-2 Killing tensor field 
on the quotient space with the effective metric.

We have demonstrated the dynamical properties of the Hopf loop strings 
in the Kerr-AdS background with equal angular momenta. 
Because of high symmetry of the metric, 
the dynamics is determined by radial motion of geodesic particle 
in the quotient space. 
By using the effective potential for the radial motion, 
we have shown that there exist bounded orbits and unbounded orbits 
in the quotient space, 
and there exist the innermost or outermost stable orbit. 
The motions of Hopf loops are driven by three forces: tension of string, 
the centrifugal force, and force of gravity \cite{Igata:2009dr},
and the orbits are determined by the competition of these forces. 
The stationary solutions are achieved by the balance of these forces. 
The existence of the bounded orbits of the Hopf loops around 
five-dimensional black holes makes us recall the geodesic particles 
around a four-dimensional black hole.

We have also shown that the Hamilton-Jacobi equation 
is completely separable for the general toroidal spiral strings 
in the five-dimensional Minkowski background if we choose appropriate coordinates. 
We have presented the general dynamical solution of the toroidal spiral strings. 
The stationary toroidal spirals have been given 
in the Ref.\cite{BlancoPillado:2007iz, Igata:2009dr}, 
and generalization in less symmetric case is discussed 
in Ref.\cite{BlancoPillado:2007iz}. 
We have presented another generalization to dynamical toroidal spirals 
in this paper.

It would be easy to replace the five-dimensional black holes with other 
target spacetimes which admit two commutable rotational Killing vector fields.  
Toroidal spiral strings in black rings are interesting targets of next study. 

\section*{Acknowledgements}
This work is supported by the Grant-in-Aid for Scientific Research No.19540305. 

%

\end{document}